\documentclass[pra,twocolumn,showpacs,amsmath,eqsecnum,amssymb,superscriptaddress,longbibliography]{revtex4-1}
                                
\usepackage{graphicx,color,upgreek,epstopdf,hyperref,breakurl} 
\usepackage{times}   

\newcommand{\tr}{\mathop{\mathrm{Tr}} \nolimits}
\newcommand{\I}{\mathrm{i}}
\newcommand{\ket}[1]{\left |#1\right\rangle}
\newcommand{\bra}[1]{\left\langle#1\right|}
\newcommand{\braket}[2]{\langle{#1}|{#2}\rangle}
\newcommand{\Eta}{\mathalpha{\textsl{H}}}  
\newcommand{\ds}{\displaystyle}  
\newcommand{\half}{\frac{1}{2}}  
\newcommand{\third}{\frac{1}{3}} 
\newcommand{\repr}{\mathrel{\widehat{=}}}  
\newcommand{\Exp}[1]{\mathrm{e}^{\mbox{\footnotesize$#1$}}}  

\begin{document}

\title{Least-bias state estimation with incomplete unbiased measurements}

\author{Jaroslav~{\v R}eh{\'a}{\v c}ek}
\affiliation{Department of Optics, Palack{\'y} University,
17.~listopadu 12, 77146 Olomouc, Czech Republic}

\author{Zden\v{e}k~Hradil}
\affiliation{Department of Optics, Palack{\'y} University,
17.~listopadu 12, 77146 Olomouc, Czech Republic}

\author{Yong~Siah~Teo}
\affiliation{Department of Optics, Palack{\'y} University,
17.~listopadu 12, 77146 Olomouc, Czech Republic}

\author{Luis~L.~S{\'a}nchez-Soto}
\affiliation{Departamento de {\'O}ptica,
Facultad de F{\'\i}sica, Universidad Complutense,
28040 Madrid, Spain}
\affiliation{Max-Planck-Institut f\"ur die Physik des Lichts,
G\"{u}nther-Scharowsky-Stra{\ss}e 1, Bau 24,
91058 Erlangen, Germany}

\author{Hui Khoon Ng}
\affiliation{Yale-NUS College, 16 College Avenue West, 
Singapore 138527, Singapore}
\affiliation{Centre for Quantum Technologies,
National University of Singapore,
3 Science Drive 2, Singapore 117543, Singapore}
\affiliation{MajuLab, CNRS-UNS-NUS-NTU International Joint Unit,
UMI 3654, Singapore}

\author{Jing~Hao~Chai}
\affiliation{Centre for Quantum Technologies,
National University of Singapore,
3 Science Drive 2, Singapore 117543, Singapore}

\author{Berthold-Georg~Englert}
\affiliation{Centre for Quantum Technologies,
National University of Singapore,
3 Science Drive 2, Singapore 117543, Singapore}
\affiliation{MajuLab, CNRS-UNS-NUS-NTU International Joint Unit,
UMI 3654, Singapore}
\affiliation{Department of Physics,
National University of Singapore,
2 Science Drive 3, Singapore 117542, Singapore}

\begin{abstract}
  Measuring incomplete sets of mutually unbiased bases constitutes a
  sensible approach to the tomography of high-dimensional quantum
  systems.  The unbiased nature of these bases optimizes the
  uncertainty hypervolume. However, imposing unbiasedness on the
  probabilities for the unmeasured bases does not generally yield the
  estimator with the largest von Neumann entropy, a popular figure of
  merit in this context.  Furthermore, this imposition typically leads
  to mock density matrices that are not even positive definite.  This
  provides a strong argument against perfunctory applications of
  linear estimation strategies. We propose to use instead the physical
  state estimators that maximize the Shannon entropy of the unmeasured
  outcomes, which quantifies our lack of knowledge fittingly and gives
  physically meaningful statistical predictions.
\end{abstract}


\maketitle

\section{Introduction}

Modern quantum technologies harness characteristic features of quantum
systems to gain performance that is otherwise unattainable through
classical means.  This progress ultimately relies on the ability to
create, manipulate, and measure quantum states. All of these tasks
require a step-by-step verification in the experimental procedures;
this is essentially the scope of quantum tomography~\cite{lnp:2004uq}.

Typically, a tomographic protocol attempts to infer the unknown
quantum state from the distinct outcomes of a collection of
measurements performed on a finite set of identical copies of the
system.  With these limited resources, the choice of optimal
measurements and the design of efficient reconstruction algorithms are
crucial.

When the measurement outcomes form an informationally complete
set~\cite{Prugovecki:1977fk,Busch:1989kx,Ariano:2004kx,
  Flammia:2005fk,Weigert:2006zr}, the data obtained contain maximal
information and a unique state estimator can be
inferred. Unfortunately, as we probe more intricate quantum
systems~\cite{Monz:2011lq,Yao:2012hc}, such an informationally
complete set of measurements becomes extremely difficult to implement.
In addition, the complete knowledge of the quantum state of a system
with many degrees of freedom is usually not needed, as one could very
well be interested in a few parameters only, such as the fidelity with
respect a target state or a measure of entanglement.

We are thus inevitably led to the consideration of alternative
techniques. A promising class of new protocols are explicitly
optimized for particular kinds of states. This includes states with
low rank~\cite{Gross:2010dq,Flammia:2012if,Guta:2012bl}, such as
matrix product states \cite{Cramer:2010oq,Baumgratz:2013fq}, or
multi-scale entanglement renormalization ansatz
states~\cite{Landon-Cardinal:2012mb}. The specific but pertinent
example of permutation invariance was also
examined~\cite{Ariano:2003qf,Toth:2010dq,Klimov:2013ya,Moroder:2012bs}.

In the same spirit, several algorithms for estimating quantum states
from incomplete measurement data have been reported
recently~\cite{Teo:2011mn,Teo:2012nb,Teo:2012ve,Goncalves:2013uz}. Here,
we revisit the problem in the context of mutually unbiased (MU)
measurements, which are known to be optimal for state
reconstruction~\cite{Wootters:1989qf} if a complete set of bases is to
be measured.  At the fundamental level, mutually unbiased bases (MUB)
are part of the mathematical framework for an explicit formalism of
the complementarity principle.  The idea emerged in the pioneering
work of Schwinger~\cite{Schwinger:1960dq}, and has since been
integrated into the foundation of quantum theory: apart from their
role in tomography, MUB are instrumental in addressing a number of
enthralling questions~\cite{Durt:2010cr}.

A first, self-suggesting, if naive, approach could be to assume a
uniform distribution for the outcomes of the unmeasured bases from a
set of MUB, as this seems to be consistent with the very physical
nature of MUB, which minimize the statistical uncertainty
hypervolumes.  However, as we show here, this is often incompatible
with the positivity required by quantum mechanics, and even when the
resulting estimator is physical, it is usually not the estimator with
the largest von Neumann entropy~\cite{Teo:2011mn}.

The bases of eigenstates of complementary observables are called
\emph{unbiased} because we cannot predict at all the outcome of a
projective measurement in one basis if the system is prepared in a
state from another basis---all outcomes are equally probable, there
is no preference in our prediction.  This notion of being unbiased
must not be confused with other uses of the adjective.  In fact, there
are many different meanings and connotations associated with ``bias'':
statistical bias, bias in a sample, cultural bias, and media bias are
but a few uses of the word.

In the context of state estimation, \emph{statistical bias} is of some
importance; an estimator has a statistical bias if its average over
all thinkable compositions of a finite sample of measurement results
deviates from the true value.  This will be of no concern here because
we shall take for granted that the measured sample is so large that
statistical noise in the data can be safely ignored.  Also, we do not
have to worry about a \emph{biased sample}, a common problem when
polls are taken.

The simple ``estimators of unbiased linear inversion,'' which we shall
introduce in Sec.~\ref{sec:2C}, are unbiased in the sense that they
assume equal probability for all outcomes of the unmeasured bases from
a set of MUB; they are, however, estimators with a statistical bias
(for data from a finite sample), a property that cannot be
avoided~\cite{Hradil:1999ij,Cousins:1995fr} if one insists, as we do,
that all permissible estimators are physical---they must be
\textsl{bona fide} density operators. As noted in the preceding
paragraph, the statistical bias is irrelevant in the current context,
and it is worth recalling that, despite the negative connotation of
``bias,'' a statistical bias is not only harmless, but can be rather
beneficial (see Jaynes's discussion in sections 17.2 and 17.3 of
Ref.~\cite{Jaynes:2003ee}).

Regarding \emph{cultural bias} and \emph{media bias}, we are confident
that we do not suffer from the former and hope that we will not be a
victim of the latter.  May Bias, Melampus's brother and Pero's and
Iphianassa's husband (consecutively), guard this work against biased
judgment.

The plan of this paper is as follows.  After introducing background
material and notational conventions in Sec.~\ref{sec:background}, we
illustrate various aspects of the said self-suggesting approach in
Sec.~\ref{sec:ULINexamples}, with particular emphasis on unphysical
properties of the resulting estimators. We conclude that the
``estimators of unbiased linear inversion'' are often
unphysical---they are \emph{not} estimators.  Then, in
Sec.~\ref{sec:least-bias}, we introduce a natural alternative that
keeps these estimators whenever they are physical and, when they are
not, replaces them with physical estimators.  Rather than being
unbiased about the probabilities of unmeasured bases, the physical
estimators are minimally biased.  In this alternative approach, we
maximize the Shannon entropy with due attention to the physical
constraints and so minimize the bias.  This entails a simple
optimization algorithm over the state space, in which a single
equation is iterated.  In Sec.~\ref{sec:LBcomp}, we consider
alternatives to maximizing the Shannon entropy.  Some of them minimize
the bias with respect to another criterion, others are characterized
in a different way.  Finally, we offer a summary and conclusions in
Sec.~\ref{sec:con}.

\section{Basic matters}
\label{sec:background}

\subsection{Mutually unbiased bases}
\label{sec:2A}

The state of a $d$-dimensional quantum system is specified by a
positive semidefinite, unit-trace density operator
$\varrho$. Carefully note that a different symbol $\rho$ is reserved
for its $d \times d$ matrix representation, which requires
${d^{2} -1}$ independent real parameters for its complete
characterization. If a von Neumann maximal test is chosen to fix
${d-1}$ of these parameters, then a total of ${d + 1}$ tests is
necessary to reconstruct the state.  This strategy is optimal when the
bases in which the measurements are carried out are ``as different as
possible;'' that is, when these bases are MU~\cite{Durt:2010cr}.

Throughout this paper, we take the dimension $d$ to be a prime or a
prime-power integer.  Then, a maximal number of ${d+1}$
MUB~\cite{Ivanovic:1981ly} exist and can be explicitly
constructed~\cite{Bandyopadhyay:2002db,Klappenecker:2003fk,Lawrence:2004xj,
  Pittenger:2004vn,Parthasarathy:2004kx, Wocjian:2005pw,
  Durt:2005ys,Klimov:2005oq, Klimov:2007bh,Klimov:2009bk}.  We denote
by $| \psi_{\alpha k}\rangle $ the $k$th ket in the $\alpha$th basis of
the set of MUB; here and below, Greek indices ($\alpha$, $\beta$,
\ldots) label the $d+1$ distinct MUB, whereas Latin indices ($k$, $l$,
\ldots) label the $d$ outcomes in each basis.

We define MU projectors as
$\Pi_{\alpha k} = | \psi_{\alpha k} \rangle \langle \psi_{\alpha k}|$.
Any two MU projectors satisfy the trace relation
\begin{equation}
  \label{MUBdef}
  \tr ( \Pi_{\alpha k} \, \Pi_{\beta \ell} )  =
  \delta_{\alpha \beta} \delta_{k \ell} +
  \frac{1}{d} (1-\delta_{\alpha \beta} )
\end{equation}
that states their orthonormality for ${\alpha=\beta}$ and their mutual
unbiasedness for ${\alpha\neq\beta}$.  Besides, these projectors
constitute a complete set of projective measurements for each
$\alpha$,
\begin{equation}
  \label{eq:1}
  \sum_{k=0}^{d-1}\, \Pi_{\alpha k} = \openone \, .
\end{equation}

To facilitate the discussion of the physics behind incomplete MUB
tomography, the number of copies of the system probed by the
measurement apparatus is taken sufficiently large, so statistical
fluctuations in the measurement data are negligible.

The problem we are studying here is, therefore, not one of state
estimation \textsl{sensu stricto}, where the estimation of the
probabilities from observed relative frequencies is the central theme.
Rather, we are dealing with the problem of converting the
probabilities into a statistical operator, which requires a deliberate
choice when the tomography is incomplete.

\subsection{Complete tomography with MUB}
\label{sec:2B}

The ${d^2+d}$ measured probabilities
\begin{equation}
\label{eq:bm-0}
p_{\alpha k} = \tr (\Pi_{\alpha k} \,\varrho)\,,
\end{equation}
which obey the ${d+1}$ constraints ${\sum_k p_{\alpha k}=1}$,
fully characterize  the density operator $\varrho$ of the system.
First, if only one basis is measured---the $\alpha$th basis, say---our
state estimator is 
\begin{equation}
\label{eq:bm-1}
\widehat{\varrho}_{\alpha}=\sum_{k=0}^{d-1}p_{\alpha k}\Pi_{\alpha k}\,.
\end{equation}
It is the most natural estimator, inasmuch as we estimate the
probabilities of the unmeasured bases in the unbiased manner of
${\widehat{p}_{\beta l}=\tr(\Pi_{\beta l}\,\widehat{\varrho}_{\alpha})=1/d}$
for ${\beta\neq\alpha}$; this is most adequate for MUB if we are
mindful of Laplace's advice to assign equal probabilities to
alternatives about which we have no information \cite{Laplace:1774ta}.
Second, after measuring all ${d+1}$ bases, we know the density
operator completely,
\begin{equation}
  \label{eq:bm-2}
  \varrho=\sum_{\alpha=1}^{d+1}\widehat{\varrho}_{\alpha} -
  \openone = \frac{1}{d} \openone + 
  \sum_{\alpha=1}^{d+1} \left  ( 
    \widehat{\varrho}_{\alpha} - \frac{1}{d} \openone \right ) \, ;
\end{equation}
alternatively, we write
\begin{equation}
  \label{eq:bm-2'}
  \varrho =\frac{1}{d} \openone +
  \sum_{\alpha=1}^{d+1} \sum_{k=0}^{d-1} 
  w_{\alpha k} \, \Pi_{\alpha k} \, , 
  \quad \mbox{with}\quad
  w_{\alpha k}  = p_{\alpha k}  - \frac{1}{d} \, .
\end{equation}
While the mappings ${\varrho\mapsto p_{\alpha k}}$ and
${p_{\alpha k} \mapsto\widehat{\varrho}_{\alpha}}$ are linear, the mappings
${p_{\alpha k} \mapsto\varrho}$ in Eqs.~\eqref{eq:bm-2} or \eqref{eq:bm-2'}
appear to be affine.  In fact, these are linear mappings too, since we
can make use of identities such as
\begin{eqnarray}
  \label{eq:bm-3}
  {\openone} & = & {\frac{1}{d+1}  
  \sum_{\alpha=1}^{d+1}  \sum_{j,k=0}^{d-1} p_{\alpha j} \Pi_{\alpha k}}  \, ,
  \nonumber \\ [1ex]
  {w_{\alpha k}} & = & {\sum_{j=0}^{d-1}p_{\alpha j}  
\left  ( \delta_{jk}-\frac{1}{d} \right )} \, , 
\end{eqnarray}
if we wish.

Here is a caveat. The linear map
${p_{\alpha k} \mapsto \widehat{\varrho}_{\alpha}}$ in \eqref{eq:bm-2}
for a single basis continues to yield physical estimators if one
replaces the probabilities $p_{\alpha k}$ by the corresponding
relative frequencies from an experiment with a finite sample.  The
resulting single-basis estimators have no statistical bias.  For the
full tomography, however, one cannot replace the probabilities in
\eqref{eq:bm-2'} by relative frequencies.  If one does, one gets mock
estimators that have no statistical bias but are unphysical, whereas
physical estimators are statistically biased since they require a
suitable nonlinear mapping from the relative frequencies to estimated
probabilities.  See Refs.~\cite{Schwemmer:2015bs,Shang:2014kb} for
various aspects of these matters.

The second version of Eq.~\eqref{eq:bm-2} exploits a well-known important
geometrical property of density operators: the difference between any
$\varrho$ and the completely mixed state $\frac{1}{d}\openone$ is a
traceless Hermitian operator, and these operators are elements of a
${d^2-1}$ real vector space.  The sum over $\alpha$ in
Eq.~\eqref{eq:bm-2} or \eqref{eq:bm-2'} is a sum over hyperplanes in
this vector space; there are ${d+1}$ of them, each
${(d-1)}$-dimensional since ${\sum_kw_{\alpha k}=0}$.  The hyperplanes
are orthogonal in the sense of
\begin{equation}
\label{eq:bm-3a}
  \tr \left ( \sum_{k=0}^{d-1} w_{\alpha k} \Pi_{\alpha k}
 \sum_{l=0}^{d-1} w_{\beta l}^{\prime} \Pi_{\beta l} \right  )=0 \, ,
\quad
\mbox{for ${\alpha\neq\beta}$}\,,
\end{equation}
and in
\begin{eqnarray}
\label{eq:bm-3b}
  \tr(\varrho\,\varrho ^{\prime} )& = & \frac{1}{d}+
\sum_{\alpha=1}^{d+1} \sum_{k=0}^{d-1} 
w_{\alpha k} w_{\alpha k}^{\prime} \,,
\nonumber\\
\tr [ (\varrho-\varrho ^{\prime})^2 ]& = &
\sum_{\alpha=1}^{d+1}\sum_{k=0}^{d-1}
( w_{\alpha k}-w_{\alpha k}^{\prime} )^2
\end{eqnarray}
we recognize the Euclidean metric of the vector space.

\subsection{Linear inversion for incomplete MUB tomography}
\label{sec:2C}

The single-basis estimator in Eq.~\eqref{eq:bm-1} is obtained from the
right-hand side of Eq.~\eqref{eq:bm-2} by setting
$\widehat{\varrho}_{\beta} \mapsto\frac{1}{d} \openone$ for the basis
indices $\beta$ that are different from the privileged index $\alpha$.
Now, if an intermediate number of bases from the set of MUB has been
measured---say, the first $M$ bases with ${1<M<d+1}$---the
unbiasedness of the bases and the linearity just noted suggest to set
$\widehat{\varrho}_{\beta} \mapsto\frac{1}{d}\openone$ for the ${d+1-M}$
unmeasured bases, which yields the $M$th estimator of \emph{unbiased
  linear inversion} (ULIN),
\begin{eqnarray}
\label{eq:bm-4}
  \widehat{\varrho}^{(M)}_\textsc{ulin} & = &
\frac{1}{d}\openone  + \sum_{\alpha=1}^{M} 
\left ( \widehat{\varrho}_{\alpha}    - 
\frac{1}{d}\openone \right ) =   
 \sum_{\alpha=1}^{M}\widehat{\varrho}_{\alpha} -
\frac{M-1}{d}\openone   \nonumber\\ 
&=&
\frac{1}{d}\openone + 
\sum_{\alpha=1}^M \sum_{k=0}^{d-1} 
w_{\alpha k}\,\Pi_{\alpha k} \, .
\end{eqnarray}
We can also regard the truncation of the $\alpha$ summation as
effected by a replacement of the coefficients $w_{\alpha k}$ in
Eq.~\eqref{eq:bm-2'} in accordance with
\begin{equation}
  \label{eq:bm-4'}
  w_{\alpha k} \mapsto \widehat{w}_{\alpha k}^{(M)} =  
   \begin{cases}
    p_{\alpha k} - \dfrac{1}{d}  & \mbox{for}
  \quad 1\leq\alpha\leq   M\,,\\ 
   0 & \mbox{for}\quad \alpha >M \, .
  \end{cases}
\end{equation}

Yet, irrespective of how we regard the map of the measured
probabilities $p_{\alpha k}$ with ${\alpha\leq M}$ onto the ULIN
estimator $\widehat{\varrho}^{(M)}_\textsc{ulin}$, the map simply
amounts to estimating the probabilities of the unmeasured bases in the
unbiased manner---faithful to Laplace's advice, so to say:
\begin{equation}
\label{eq:bm-5}
    \widehat{p}_{\beta l}
   =\tr ( \Pi_{\beta l} \,\widehat{\varrho}^{(M)}_\textsc{ulin} )
   =\frac{1}{d}  \quad\mbox{for ${\beta>M}$}\,.
\end{equation}
All other estimators are characterized by the non-zero operator they
assign to the unmeasured part in
\begin{equation}
 \label{eq:bm-5a}
  \widehat{\varrho}^{(M)}_{\text{other}} = 
  \widehat{\varrho}^{(M)}_\textsc{ulin}
  +\left ( \widehat{\varrho}
    -\frac{1}{d}\openone \right )^{(M)}_{\text{unmeasured}}\,, 
\end{equation}
where
\begin{equation}
\label{eq:bm-5b}
  \left (\widehat{\varrho} -
    \frac{1}{d}\openone \right )^{(M)}_{\text{unmeasured}}=
  \sum_{\beta=M+1}^{d+1} \sum_{l=0}^{d-1}
  \widehat{w}_{\beta l, \text{other}}\,\Pi_{\beta l}
\end{equation}
is the estimator for the contribution of the unmeasured bases to the
sums in Eqs.~\eqref{eq:bm-2} and \eqref{eq:bm-2'}.

Before proceeding, six comments are in order.  First, the map
${\varrho \mapsto \widehat{\varrho}^{(M)}_{\textsc{ulin}}}$ projects
${\varrho-\frac{1}{d}\openone}$ onto the span of the first $M$
hyperplanes in the vector space discussed above; applying the map a
second time has no effect.  This projection property implies
\begin{eqnarray}
 \label{eq:bm-5c}
  \tr ( \widehat{\varrho}^{(M)}_{\textsc{ulin}}\,\varrho^{\prime} ) &=&
  \tr ( \varrho\,\widehat{\varrho'}^{(M)}_{\textsc{ulin}} ) =
  \tr (\widehat{\varrho}^{(M)}_{\textsc{ulin}}\,
   \widehat{\varrho'}^{(M)}_{\textsc{ulin}} ) \nonumber \\
  &=& \frac{1}{d} +
   \sum_{\alpha=1}^M \sum_{k=0}^{d-1} 
w_{\alpha k} w_{\alpha k}^{\prime} \, , 
\end{eqnarray}
for any two density operators $\varrho$ and $\varrho^{\prime}$.

Second, we note that, since Eq.~\eqref{eq:bm-4'} involves only the
measured bases, it is also well defined for systems where $M$ MUB
exist, but this set cannot be extended to $d+1$ bases. This is the
case, e.g., for ${d=6}$ and ${M=3}$ where only sets of at most three
MUB have been found thus far~\cite{Zauner:2011nu,Butterley:2007fe,
  Bengtsson:2007lb,Brierley:2008ja,Brierley:2009mw,Jaming:2009ck}.  
There are also non-extendable sets for prime-power
dimensions~\cite{Mandayam:2014sa}, although ULIN estimators for these
sets  suffer from the same problems as the ULIN estimators for subsets
of full  sets (see Sec.~\ref{sec:ULINexamples}): they are often unphysical.

Also, whether the unmeasured bases are pairwise unbiased among
themselves, or unbiased with some of the measured bases, is not
important for the maximum property of the following third comment, nor
for the physical least-bias estimators of Secs.~\ref{sec:LBphilo} and
\ref{sec:LBcomp-1}.  If one wishes, one can choose the unmeasured
bases such that they have largest average distance (see
Ref.~\cite{Raynal:2011xc}), so that they resemble a set of MUB as best
as they can.

 Third, it is also worth noting that, of all estimators consistent with
the measured probabilities, the ULIN estimator
$\widehat{\varrho}^{(M)}_\textsc{ulin}$ maximizes the Shannon entropy
\begin{equation}
  \label{eq:bm-6}
  \Eta^{(M)}(\varrho) = - \sum^{d+1}_{\beta=M+1} \sum^{d-1}_{l=0}
  {p}_{\beta l} \ln p_{\beta l} 
\end{equation}
associated with the unmeasured probabilities, which must be calculated
according to Eq.~\eqref{eq:bm-0}. The ULIN estimator has this property
by construction, since it gives the uniform estimates of
Eq.~\eqref{eq:bm-5} for the $\widehat{p}_{\beta l}$s.

For simplicity, we are taking some liberties with the Shannon entropy
in \eqref{eq:bm-6}: the probabilities are normalized to unit sum for
each $\beta$, rather than to total unit sum and  we are using the natural
rather than the binary logarithm. This betrayal of the pure doctrine
is of no consequence, however. 

Fourth, irrespective of the unbiased way of estimating the
$\widehat{p}_{\beta l}$s, the ULIN estimator has a statistical bias;
recall the pertinent remarks in the Introduction.

Fifth, as we shall see, after measuring $M$ bases, we actually have
some information about the ${d+1-M}$ unmeasured bases.  We are not
really faithful to Laplace's advice as long as we do not account for
this information properly. For ${M=1}$ we recover the
estimator~(\ref{eq:bm-1}) with ${\alpha=1}$, whereas
${\widehat{\varrho}^{(d+1)}_\textsc{ulin}=\varrho}$ is the
full-tomography case of ${M=d+1}$ in Eq.~\eqref{eq:bm-2}.  The
reliability of $\widehat{\varrho}^{(M)}_\textsc{ulin}$ for these
limiting values of $M$ is, however, misleading, and so is the
self-suggesting probability estimation~(\ref{eq:bm-5}).  As we shall
see---and this is one main objective of this work---the ULIN
estimator is not assuredly physical for intermediate $M$ values,
except in the qubit situation of ${d=2}$.  That is,
$\widehat{\varrho}^{(M)}_\textsc{ulin} \ngeq 0$ is possible.  The
correct implementation of Laplace's advice is not achieved by the
simple-minded estimates in \eqref{eq:bm-5}.

Sixth, there is no other linear map ${p_{\alpha k} \mapsto 
\widehat{\varrho}^{(M)}_\textsc{lin}}$ that could be used instead of
${p_{\alpha k} \mapsto\widehat{\varrho}^{(M)}_\textsc{ulin}}$ if we
require for consistency, as we must, that
${p_{\alpha k}=\tr(\widehat{\varrho}^{(M)}_\textsc{lin}\, 
\Pi_{\alpha k})}$ for the $M$ bases that have been measured.  
There are, of course, legitimate state estimators
$\widehat{\varrho}^{(M)}$  that are consistent with the known 
probabilities for bases ${\alpha=1,2,\dots,M}$---namely the 
physical ``other'' estimators of Eq.~\eqref{eq:bm-5a}---and these 
estimators make up a convex set; correspondingly, there is a convex 
set of estimated probabilities $\widehat{p}_{\beta l}$ for 
${\beta=M+1,\dots,d+1}$.  The ULIN estimator of Eq.~\eqref{eq:bm-4} 
and the unbiasedly estimated probabilities ${\widehat{p}_{\beta l}=1/d}$ 
may or may not be in the respective sets.  When they are outside, we 
have to choose the best state estimator in accordance with a suitable 
optimality criterion. More about this in Sec.~\ref{sec:least-bias}.

\section{Aspects of incomplete MUB tomography}
\label{sec:ULINexamples}

\subsection{ULIN estimators for a single qubit (${d=2}$)}
\label{sec:qubit}

As a first illustration, let us take the elementary example of a
single qubit ($d=2$).  The three MUB here are just the eigenstates of
the standard Pauli operators $\sigma_x$, $\sigma_y$, and $\sigma_z$,
so that ${\Pi_{10}=\half( \openone+\sigma_x)\equiv\Pi_{x+}}$, \dots,
${\Pi_{31}=\half(  \openone -\sigma_z) \equiv \Pi_{z-}}$ and
\begin{equation}
  \label{eq:bit-1}
  \varrho=\widehat{\varrho}^{(3)}_\textsc{ulin}= \frac{1}{2} 
  ( \openone+ s_x \sigma_x+s_y \sigma_y+s_z \sigma_z) \, ,
\end{equation}
where $\mathbf{s}=(s_x,s_y,s_z)$ is the three-dimensional Bloch vector
with, e.g., ${s_x=\tr(\varrho\,\sigma_x)=\half(p_{x+}-p_{x-})}$.  The
eigenvalues of $\varrho$ are $\half(1\pm|\mathbf{s}|)$, and the von
Neumann entropy
\begin{equation}
\label{eq:bit-2}
  S(\varrho)=-\tr (\varrho\ln\varrho)
\end{equation}
grows monotonously with shrinking Bloch-vector length~$|s|$.

When a single basis is measured (${M=1}$), say that of $\sigma_x$,
only $s_x$ is known, and we have
\begin{equation}\label{eq:bit-3}
  \widehat{\varrho}^{(1)}_\textsc{ulin}=\frac{1}{2}
  ( \openone+  s_x \sigma_x)\,.
\end{equation}
Clearly, this physical estimator maximizes the von Neumann entropy,
since ${\widehat{p}_{y\pm}=\widehat{p}_{z\pm}=\half}$ give the shortest
Bloch vector consistent with the known probabilities
${p_{x\pm}=\half(1\pm s_x)}$.
Likewise, the ULIN estimator for two measured bases,
\begin{equation}\label{eq:bit-4}
  \widehat{\varrho}^{(2)}_\textsc{ulin}=\frac{1}{2}
  ( \openone+  s_x \sigma_x+s_y \sigma_y)\,,
\end{equation}
which gives the estimated probabilities
${\widehat{p}_{z_+}=\widehat{p}_{z_-}=\half}$ for the unmeasured
$\sigma_z$, maximizes
the von Neumann entropy regardless of the values for the observed
probabilities $p_{x_\pm}$ and~$p_{y_\pm}$.

These properties of the ULIN estimators for a qubit are consequences
of the simple spherical geometry of the Bloch ball and the
orthogonality of the planes corresponding to the distinct single-qubit
MU Pauli observables. Such lines of argument were employed for
justifying the optimality of MUB tomography in higher
dimensions~\cite{Wootters:1989qf}: when picturing the results obtained
with a finite number of copies using ``fuzzy'' hyperplanes, their
mutual orthogonality makes the uncertainty hypervolume particularly
small.

So, for a qubit, the unbiased unmeasured probabilities of the ULIN
estimator, as quantified by the Shannon entropy, optimize (maximize)
the von Neumann entropy, much like the unbiased MU observables
optimize (minimize) statistical uncertainty.  These two properties
complement each other.  In fact, for any ${d\geq2}$ and $M$, if
$\varrho$ is equal to a projector $\Pi_{\alpha k}$ of one of the
measured bases, then the ULIN estimator is the state itself according
to the defining relation~\eqref{MUBdef}.  In view of all this
evidence, one might expect this mutual compatibility to extend to
higher dimensions---and so arrive at the ULIN estimator of
Eq.~\eqref{eq:bm-4}.  However, we shall see that this somewhat naive
approach fails for ${d>2}$.

\subsection{ULIN estimators for a single qutrit (${d=3}$)}
\label{sec:qutrit}

For the ${d=3}$ case of a qutrit, we use this set of four MUB:
\begin{eqnarray}
 \label{eq:trit-1}
 \mathcal{B}_1 &\repr &\frac{1}{\sqrt{3}}
 \begin{pmatrix}
   1 &  1  &  1  \\
   1 & q^2 &  q  \\
   1 &  q  & q^2
 \end{pmatrix}, \quad  
 \mathcal{B}_2\repr\frac{1}{\sqrt{3}}
 \begin{pmatrix}
   1 &  1  &  1  \\
   1 & q^2 &  q  \\
   q & q^2 &  1
 \end{pmatrix}, \nonumber\\
 \mathcal{B}_3&\repr&\frac{1}{\sqrt{3}}
 \begin{pmatrix}
   1  &  1  &  1  \\
   1  & q^2 &  q  \\
  q^2 &  1  &  q
 \end{pmatrix}, \quad 
 \mathcal{B}_4\repr
 \begin{pmatrix}
  1 & 0 & 0 \\
  0 & 1 & 0 \\
  0 & 0 & 1
 \end{pmatrix},
\end{eqnarray}
where the columns are the probability amplitudes of the basis kets
with reference to the fourth basis, and ${q=\Exp{\I 2\pi/3}}$ is the
basic cubic root of unity.  The $3\times3$ matrices that represent the
respective single-basis estimators of Eq.~\eqref{eq:bm-1} are
\begin{eqnarray}
\label{eq:trit-2}
  \widehat{\rho}_1& = &\third
  \begin{pmatrix}
    1     &  z_1   & z_1^*  \\
    z_1^* &  1     & z_1   \\
    z_1   &  z_1^* & 1
  \end{pmatrix}, \qquad \quad \;
  \widehat{\rho}_2=\third
  \begin{pmatrix}
    1     &  z_2    & q^2z_2^*  \\
    z_2^* &  1      & q^2z_2   \\
    qz_2  &  qz_2^* & 1
  \end{pmatrix} \, ,\nonumber\\
  \widehat{\rho}_3& = & 
  \third
  \begin{pmatrix}
    1      &  z_3      & qz_3^*  \\
    z_3^*  &  1        & qz_3   \\
    q^2z_3 &  q^2z_3^* & 1
  \end{pmatrix} \, ,
 \quad
\widehat{\rho}_4=
  \begin{pmatrix}
  p_{40}  &  0  &  0  \\
  0  &  p_{41}  &  0  \\
  0  &  0  &  p_{42}
  \end{pmatrix},\nonumber\\
\end{eqnarray}
where
\begin{equation}
\label{eq:trit-3}
z_{\alpha}=\sum_{k=0}^2q^kp_{\alpha k} = 
\tr ( \varrho\,Z_{\alpha} )  \, ,
\quad \mbox{with} \quad 
Z_{\alpha}=\sum_{k=0}^2q^k\Pi_{\alpha k} \, .
\end{equation}
We have
${p_{\alpha k}=\third+w_{\alpha k} =
  \third(1+q^{-k}z_{\alpha}+q^kz_{\alpha}^*)}$
for the probabilities in terms of the $z_{\alpha}$s, and the unitary
$3\times3$ matrices for the pairwise complementary observables
$Z_{\alpha}$ are
\begin{eqnarray}
\label{eq:trit-4}
  Z_{\alpha} & \repr &
  \begin{pmatrix} 
0 & 0 & q^{1-\alpha}\\
1 & 0 & 0 \\
0 & q^{\alpha-1} & 0
\end{pmatrix} \, ,
 \quad \alpha=1,2,3 \,, 
 \nonumber\\
Z_4 & \repr & 
 \begin{pmatrix}
1&0&0 \\
0&q&0 \\
0&0&q^2
\end{pmatrix} \, .
\end{eqnarray}
We note that
${\widehat{\rho}_{\alpha}=
  \third(\openone+z_{\alpha}^*Z_{\alpha}+z_{\alpha}Z_{\alpha}^{\dagger})}$,
and the replacement
${\widehat{\varrho}_{\beta} \mapsto\third\openone}$ for an unmeasured
basis is here simply implemented by ${z_{\beta} \mapsto0}$.

Consider now a qutrit state for which the density operator is an
incoherent mixture of two projectors, one each from the first and the second
bases,
\begin{equation}
\label{eq:trit-5}
  \varrho=\lambda_1\Pi_{1j}+\lambda_2\Pi_{2k} \, ,
\end{equation}
with $0 \leq \lambda_1=1-\lambda_2 \leq 1$.  As an immediate
consequence of
${\tr(\Pi_{\alpha k}\,Z_{\beta})=\delta_{\alpha\beta}q^k}$, we have
here ${z_1=\lambda_1q^j}$, ${z_2=\lambda_2q^k}$, and 
${z_3=z_4=0}$. It follows that ${\widehat{\varrho}_{\textsc{ulin}}^{(2)} =
\widehat{\varrho}_{\textsc{ulin}}^{(3)}
  =\widehat{\varrho}_{\textsc{ulin}}^{(4)}=\varrho}$,
so that all ULIN estimators are genuine density operators for this
particular~$\varrho$.

Matters are quite different for
\begin{equation}
\label{eq:trit-6}
   \rho = \half
 \begin{pmatrix}
1&-1&0\\
-1&1&0\\
0&0&0
\end{pmatrix} \, ,
\end{equation}
for which ${z_1=z_2=z_3=-\half}$ and ${z_4=-\half q^2}$; all
$\widehat{\varrho}_{\alpha}$s have eigenvalues $0,\half,\half$.  While
${\widehat{\varrho}_{\textsc{ulin}}^{(1)}=\widehat{\varrho}_1}$ and
${\widehat{\varrho}_{\textsc{ulin}}^{(4)}=\varrho}$ are density
operators, as they are always, here
$\widehat{\varrho}_{\textsc{ulin}}^{(2)}$ and
$\widehat{\varrho}_{\textsc{ulin}}^{(3)}$ are not acceptable
estimators; they have negative determinants of $-\frac{1}{27}$ and
$-\frac{5}{108}$, respectively.

The last example illustrates the statement in Sec.~\ref{sec:2C} that
``the ULIN estimator is not assuredly physical for intermediate $M$
values,'' here for ${1<M<d+1=4}$.  Analogous examples can be
constructed for higher dimensions; see Sec.~\ref{sec:qudit}.
Regarding the other assertion in Sec.~\ref{sec:2C}--- that there is
no consistent ``linear map
${p_{\alpha k} \mapsto\widehat{\varrho}^{(M)}_\textsc{lin}}$ that could be
used instead of
${p_{\alpha k} \mapsto\widehat{\varrho}^{(M)}_\textsc{ulin}}$''---we
shall now argue that one gets a contradiction when assuming that there
is such a map.  For this purpose, we look specifically at the case of
${M=d=3}$, but the argument can clearly be modified for other
dimensions ${d>2}$ and other intermediate $M$ values.

Suppose we have measured the probabilities of bases $1$, $2$, and $3$
but lack data for basis $4$.  Then we know the values of $z_1$, $z_2$,
and $z_3$, which determine all off-diagonal elements in
$\widehat{\rho}_{\textsc{lin}}^{(3)}$, and we need a linear (or
affine) map $(z_1,z_2,z_3) \mapsto\hat{z}_4^{(3)}$ for the estimation of
the diagonal matrix elements.  As an immediate consequence of the
linearity, we note that convex sums of $(z_1,z_2,z_3)$ values yield
respective convex sums of $\hat{z}_4^{(3)}$ values,
\begin{eqnarray}
\label{eq:trit-7}
& \hat{z}_4^{(3)}  (\lambda z_1+\lambda^{\prime} z_1^{\prime} ,
\lambda z_2 +\lambda^{\prime} z_2^{\prime}, 
\lambda z_3 + \lambda^{\prime} z_3^{\prime} ) & 
\nonumber \\
& = 
  \lambda\hat{z}_4^{(3)}(z_1,z_2,z_3) +
 \lambda^{\prime}\hat{z}_4^{(3)}
 (z_1^{\prime},   z_2^{\prime},z_3^{\prime} ) \, , & 
\end{eqnarray}
for ${0\leq \lambda=1-\lambda'\leq1}$.
Further, for ${z_1=z_2=z_3=u}$ with ${0<|u|\leq\half}$, we have
\begin{eqnarray}
\label{eq:trit-8}
\widehat{\rho}_{\textsc{lin}}^{(3)} & = &
\begin{pmatrix}
\cdot & u & 0 \\ 
u^\ast & \cdot & 0 \\ 
0 & 0 &\cdot
\end{pmatrix}
\\ \nonumber 
&=&
 \lambda
\begin{pmatrix}
\cdot&\half\Exp{\I\varphi}&0 \\
\half\Exp{-\I\varphi}&\cdot&0 \\
0&0&\cdot
\end{pmatrix}
+\lambda^{\prime} 
\begin{pmatrix}
\cdot&-\half\Exp{\I\varphi}&0 \\
-\half\Exp{-\I\varphi}&\cdot&0 \\
 0&0&\cdot
\end{pmatrix} \, ,
\end{eqnarray}
with ${\Exp{\I\varphi}=u/|u|}$ and ${\lambda=\half+|u|}$,
and the dots stand in for the yet-unknown diagonal entries.
In the convex sum, both matrices must have $p_{40}=p_{41}=\half$ 
and $p_{42}=0$ on the diagonal, which implies the same values for this
$\widehat{\rho}_{\textsc{lin}}^{(3)}$.
We conclude that
\begin{equation}
\label{eq:trit-9}
  \hat{z}_4^{(3)}(u,u,u) = \half(1+q) = - \half q^2 \,,
\end{equation}
which, by convexity, also holds for ${u=0=\half u'+\half(-u')}$.
Analogous arguments for
\begin{equation}
\label{eq:trit-10}
 z_1= q z_2 = q^2 z_3 =u \,: 
\quad 
\widehat{\rho}_{\textsc{lin}}^{(3)} =
\begin{pmatrix}
\half&0&u^\ast \\ 
0&0&0 \\ 
u&0&\half
\end{pmatrix} \, ,
\end{equation}
and
\begin{equation}
\label{eq:trit-11}
 z_1=q^2z_2=qz_3=u\,:\quad 
\widehat{\rho}_{\textsc{lin}}^{(3)} =
\begin{pmatrix}
0 & 0 & 0 \\ 
0 &\half & u \\ 
0 &u^\ast &\half
\end{pmatrix} \, ,
\end{equation}
establish
\begin{eqnarray}
\label{eq:trit-12}
    \hat{z}_4^{(3)}(u,q^2u,qu)&=&\half(1+q^2)=-\half q \,,
  \nonumber\\
 \hat{z}_4^{(3)}(u,qu,q^2u)&=&\half(q+q^2)=-\half\,.
\end{eqnarray}
Obviously, there is a contradiction for ${z_1=z_2=z_3=0}$, as we
cannot have $-\half$ and $-\half q$ and also $-\half q^2$ for
$\hat{z}_4^{(3)}(0,0,0)$.  Case closed.

\subsection{ULIN estimators for higher dimensions}
\label{sec:qudit}

The ${d=3}$ example of Eq.~\eqref{eq:trit-6} is easily generalized to
any dimension $d$.  The ULIN estimator for ${M=d}$,
\begin{equation}
\label{eq:dit-1}
\widehat{\rho}^{(d)}_\textsc{ulin}=
\begin{pmatrix}
1/d & -1/2 & 0 & 0 &\ldots\\
-1/2 & 1/d & 0 & 0 & \ldots\\
0 & 0 & 1/d & 0  & \ldots\\
0 & 0 & 0 & 1/d  & \\
\vdots & \vdots & \vdots &  &\ddots
\end{pmatrix}\not\geq0 \,,
\end{equation}
is unphysical for all ${d>2}$ because it has the negative eigenvalue
${1/d-1/2}$.

Yet another example of unphysical ULIN estimators for ${1<M<d+1}$ is
provided by a projector to a superposition of two states from the first basis
[the $\varrho$s of Eqs.~\eqref{eq:trit-6} and \eqref{eq:dit-1} project on
superpositions of two states from the $(d+1)$th basis]; that is,
\begin{eqnarray}
\label{eq:dit-2}
\varrho & = & 
 (\ket{\psi_{10}} +\ket{\psi_{11}} ) \half 
( \bra{\psi_{10}} +\bra{\psi_{11}} )
\nonumber\\&=&\mathcal{B}_1
\begin{bmatrix}
  1/2 & 1/2 & 0 & \cdots \\
  1/2 & 1/2 & 0 & \cdots \\
   0  &  0  & 0 & \cdots \\
   \vdots & \vdots &\vdots & \ddots
\end{bmatrix}
\mathcal{B}_1^{\dagger}\,,
\end{eqnarray}
where $\mathcal{B}_1=\left ( 
\begin{array}{cccc}
\ket{\psi_{10}} & \ket{\psi_{11}} &
 \cdots & \ket{\psi_{1d-1}} \end{array} \right)$
is the row of kets from the first basis, as in Eq.~\eqref{eq:trit-1},
and $\mathcal{B}_1^{\dagger}$ is the adjoint column of bras. Note 
that we are using square parentheses to denote matrices 
 expressed with the first basis, as opposed to the usual
computational basis. 

Here,
\begin{equation}
\label{eq:dit-3}
  \widehat{\varrho}_1 = 
 \widehat{\varrho}^{(1)}_{\textsc{ulin}}
=\mathcal{B}_1
\begin{bmatrix}
  1/2 & 0 & 0 & \cdots \\
  0 & 1/2 & 0 & \cdots \\
   0  &  0  & 0 & \cdots \\
   \vdots & \vdots &\vdots & \ddots
\end{bmatrix}
\mathcal{B}_1^{\dagger}
\end{equation}
accounts for all the diagonal elements whereas, for $\alpha=2,3,\dots,d+1$,
the nonzero matrix elements of
\begin{eqnarray}
\label{eq:dit-4}
\widehat{\varrho}_{\alpha}-\frac{1}{d}\openone
&=&
\mathcal{B}_{\alpha}
\begin{bmatrix}
  w_{\alpha0} & 0 &  \cdots & 0\\
  0 & w_{\alpha1} & \cdots & 0\\
   \vdots &  &  \ddots & \vdots\\
  0 & \hdotsfor{2} & w_{\alpha\,d-1}
\end{bmatrix}
\mathcal{B}_{\alpha}^{\dagger}
\nonumber\\&=&
\mathcal{B}_1
\begin{bmatrix}
  0 & \ast & \ast & \cdots \\
  \ast & 0 & \ast & \cdots \\
  \ast & \ast & 0 & \cdots \\
   \vdots & \vdots &\vdots & \ddots
\end{bmatrix}
\mathcal{B}_1^{\dagger}  \, .
\end{eqnarray}
Here, the matrix to the $\alpha$th basis is diagonal with the $k$th
entry equal to ${w_{\alpha k}=\mathrm{Re} (\braket{\psi_{\alpha k}}{\psi_{10}}/
  \braket{\psi_{\alpha k}}{\psi_{11}})}$,
whereas the matrix to the first basis has null entries on the diagonal
and some, if not all, off-diagonal entries are nonzero (indicated by
the symbol $*$).  Then, for $M=2,3,\dots,d$, the $d\times d$ matrix in
\begin{equation}
\label{eq:dit-5}
 \widehat{\varrho}^{(M)}_{\textsc{ulin}}
=\mathcal{B}_1
\begin{bmatrix}
  1/2 & \ast  & \ast & \cdots & \ast \\
  \ast & 1/2 & \ast & \cdots & \ast \\
  \ast &  \ast  & 0 & \cdots & \ast \\
   \vdots & \vdots &\vdots & \ddots & \vdots \\
  \ast & \ast & \ast & \cdots & 0
\end{bmatrix}
\mathcal{B}_1^{\dagger}
\end{equation}
has Hermitian $2\times2$ submatrices
$\left [ \begin{array}{cc} \ast & \ast \\ 
\ast &0\end{array}\right]\not\geq0$ with
nonzero off-diagonal elements and one or two vanishing diagonal
elements.  Such a $2\times2$ matrix has a negative determinant and
cannot be positive semidefinite.  It follows that
${\widehat{\varrho}^{(M)}_{\textsc{ulin}}\not\geq0}$ is unphysical for
intermediate $M$ values.

For illustration, we take once more the qutrit MUB in
Eq.~\eqref{eq:trit-1}, for which
\begin{equation}
\label{eq:dit-6}
  \begin{array}[b]{l}\ds
    \widehat{\rho}^{(1)}_{\textsc{ulin}}=\half
    \begin{bmatrix}
1 & 0 & 0\\
0 & 1 & 0 \\
0 & 0 & 0 
\end{bmatrix} \, ,
\qquad \quad \,
\widehat{\rho}^{(2)}_{\textsc{ulin}}= \frac{1}{6}
    \begin{bmatrix}
3 & 1 & q\\
1 & 3 & q\\
q^2 & q^2 & 0 
\end{bmatrix},\\[4ex]\ds
    \widehat{\rho}^{(3)}_{\textsc{ulin}}=\frac{1}{6}
    \begin{bmatrix}
3 & 2 & -1\\
2 & 3 &-1 \\
-1 & -1 & 0
\end{bmatrix} \, ,
 \quad
    \widehat{\rho}^{(4)}_{\textsc{ulin}}=\half
    \begin{bmatrix}
1 & 1 & 0\\
1 & 1 & 0\\
0 & 0 & 0
\end{bmatrix} \, .
  \end{array}
\end{equation}
Indeed, ${\widehat{\varrho}^{(2)}_{\textsc{ulin}}}$ and
${\widehat{\varrho}^{(3)}_{\textsc{ulin}}}$ are unphysical.

In summary, then, we have seen in so many examples that ULIN
estimators can be unphysical. The unbiased estimators
${\widehat{\varrho}_{\beta}=\frac{1}{d}\openone}$ are not permissible
for ${\beta>M}$ and a nonzero ``unmeasured'' term is needed in
Eq.~\eqref{eq:bm-5a} for a physical estimator
${\widehat{\varrho}^{(M)}_{\text{other}}}$.  In this situation, the
$Md$ measured probabilities $p_{\alpha k}$ with ${\alpha \leq M}$ impose
constraints on the $(d+1-M)d$ unmeasured probabilities, which exclude
the use of ${\widehat{p}_{\beta l}=\frac{1}{d}}$ for ${\beta>M}$.  It
follows that we have some, if limited, knowledge about the observables
associated with the unmeasured bases: our best guess is \emph{not}
that all outcomes are equally likely for the future von Neumann test
of a basis not measured as yet.  Put differently, physical state
estimators for ${1<M<d+1}$ typically possess some bias in the
unmeasured probabilities, such that they deviate from the
uninformative uniform distribution.  This is the reason why we cannot
implement Laplace's advice by the naive estimates of
Eq.~\eqref{eq:bm-5}: it is simply not true that we have no information
at all about the unmeasured probabilities.

Harking back to the qubit estimators in Eqs.~\eqref{eq:bit-3} and
\eqref{eq:bit-4}, we observe that these also contain information about
the unmeasured bases because ${x^2+y^2+z^2\leq1}$ must hold for all
physical estimators, so a known value for $x$ restricts both $y$
and $z$, and known values for $x$ and $y$ bound the acceptable $z$
values.  Since ${\widehat{y}=\widehat{z}=0}$ or ${\widehat{z}=0}$,
respectively, are always permissible estimates, the intermediate ULIN
estimators work for the qubit.  This feature, however, has no analog
for ${d>2}$.

\subsection{Nonpositivity of ULIN estimators}
\label{sec:bound}

Before moving on to discussing the proper choice for the
``unmeasured'' contribution to
$\widehat{\varrho}^{(M)}_{\text{other}}$ in Eq.~\eqref{eq:bm-5a}, let
us close the subject of negative eigenvalues of intermediate ULIN
estimators by a more quantitative study.  We ask this question: what
is the most negative eigenvalue $\lambda_{\text{min}}$ that
$\widehat{\varrho}^{(M)}_{\textsc{ulin}}$ can possibly have?

For given $\varrho$, the smallest eigenvalue of
$\widehat{\varrho}^{(M)}_{\textsc{ulin}}$ can be found by minimizing
$\tr (\widehat{\varrho}^{(M)}_{\textsc{ulin}}\,\sigma)$ over all
density operators $\sigma$, and further minimization over $\varrho$
establishes
\begin{equation}
\label{eq:neg-1}
  \lambda_{\text{min}}
  =\min_{\varrho,\sigma}
   \tr (\widehat{\varrho}^{(M)}_{\textsc{ulin}}\,\sigma )
 = \min_{\varrho,\sigma}
   \tr (\varrho\,\widehat{\sigma}^{(M)}_{\textsc{ulin}} )
  \geq -\frac{M-1}{d} \,.
\end{equation}
Here, the roles of $\varrho$ and $\sigma$ are interchangeable thanks to the
symmetry noted in Eq.~\eqref{eq:bm-5c}, and the lower bound follows from
Eq.~\eqref{eq:bm-4} in conjunction with
$\tr\bigl(\widehat{\varrho}_{\alpha}\,\sigma\bigr)\geq0$ for all $\alpha$.

\begin{figure}[t]
  \centering
  \includegraphics[scale=0.95]{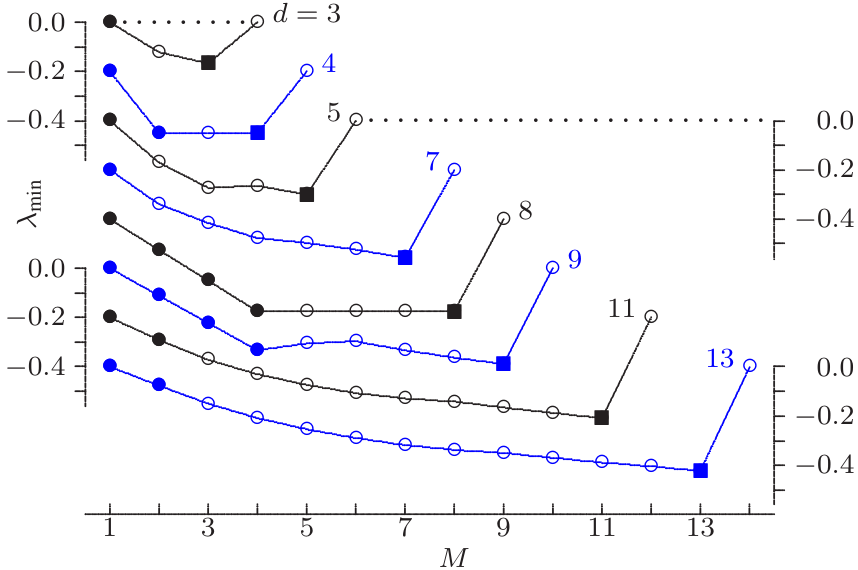}
  \caption{\label{fig:1} The most negative possible eigenvalue
    $\lambda_{\text{min}}$ of a ULIN estimator
    $\widehat{\varrho}^{(M)}_{\textsc{ulin}}$ in partial MUB
    tomography, for prime-power dimensions in the range
    ${3\leq d\leq13}$.  The successive $\lambda_{\text{min}}$ values
    of ${M=1,2,\dots,d+1}$ are connected by straight-line segments
    that guide the eye.  The curves are shifted by $-0.2$ relative to
    the preceding $d$ value. The vertical-axis labels are shown for
    ${d=3}$ and ${d=9}$ on the left, and for ${d=5}$ and ${d=13}$ on
    the right.  Each plot point represents the optimal value of
    Eq.~\eqref{eq:neg-1} over one hundred repeated numerical searches
    for solutions of the equation pair \eqref{eq:neg-2} with the
    algorithm discussed in the text.  Filled circles ($\bullet$)
    indicate values with ``$=$'' in Eq.~\eqref{eq:neg-1}, including
    ${M=1}$ always; empty circles ($\circ$) mark values with ``$>$'',
    including ${M=d+1}$ always.  For ${M=d}$, where ``$>$'' is the
    case, we have the negative eigenvalue ${\frac{1}{d}-\half}$ of the
    example in Eq.~\eqref{eq:dit-1}, shown by filled squares
    ({\tiny$\blacksquare$}).  }
\end{figure}

The equality holds in Eq.~\eqref{eq:neg-1} only if
$\sum_{\alpha=1}^M\widehat{\varrho}_{\alpha}$ is rank-deficient, and
this is the case for certain $(d,M)$ pairs but not for others; see
Fig.~\ref{fig:1}.  For example, ${\lambda_{\text{min}}=-\frac{1}{4}}$
can be established for ${(d,M)=(4,2)}$ by a simple explicit example.
Whether tight or not, the lower bound in Eq.~\eqref{eq:neg-1} tells us that
partial MUB tomography with large $d$ and small $M$ is very nearly
unbiased. For example, if only two of the MUB are measured, the lowest
minimum eigenvalue of the ULIN estimator cannot be smaller than
$-1/d$, and would approach zero for large dimensions.  Yet, even a
small negative eigenvalue renders
$\widehat{\varrho}^{(M)}_{\textsc{ulin}}$ unphysical if this
eigenvalue is different from zero within the numerical accuracy with
which it is known.

The extremal density operators $\varrho$ and $\sigma$
minimizing the traces in
Eq.~\eqref{eq:neg-1} are such that $\sigma$ projects on the smallest
eigenvalue of $\widehat{\varrho}^{(M)}_{\textsc{ulin}}$ and $\varrho$ projects
on the smallest eigenvalue of $\widehat{\sigma}^{(M)}_{\textsc{ulin}}$, so
that we have the pair of equations
\begin{equation}
\label{eq:neg-2}
 \widehat{\varrho}^{(M)}_{\textsc{ulin}}\sigma
=\tr ( \widehat{\varrho}^{(M)}_{\textsc{ulin}}\sigma ) \sigma \,,
\quad
\widehat{\sigma}^{(M)}_{\textsc{ulin}}\varrho
=\tr ( \widehat{\sigma}^{(M)}_{\textsc{ulin}}\varrho ) \varrho\,.
\end{equation}
We solve them by iteration:

\noindent\begin{tabular}{@{}p{20pt}@{}p{226pt}@{}}
\textbf{S1:} & {For the current $\varrho$, diagonalize 
$\widehat{\varrho}_{\textsc{ulin}}^{(M)}$ and set $\sigma$
to the projector onto the smallest eigenvalue.} \\[1ex]
\textbf{S2:} & {Diagonalize $\widehat{\sigma}_{\textsc{ulin}}^{(M)}$ and 
update $\varrho$ by setting it to the projector onto the smallest 
eigenvalue.} \\[1ex]
\textbf{S3:} & {Repeat steps S1 and S2 until (\ref{eq:neg-2}) holds with the
desired numerical accuracy.}
\end{tabular}

Since we are minimizing a convex function over a convex domain, the
possibility of landing at a suboptimal point on the boundary cannot be
ruled out.  Restarting the numerical search several times with
different initial states is necessary.

Figure~\ref{fig:1} shows these most negative eigenvalues for the
prime-power Hilbert-space dimensions from ${d=3}$ to ${d=13}$.  For
${d=7}$ and ${M=2}$, for instance, we find that
${\lambda_{\text{min}}\approx -0.1394}$ while the example of
Eq.~\eqref{eq:dit-2} gives a most negative eigenvalue of about
$-0.1250$.  Further, we observe the following features:

\noindent\begin{tabular}{@{}p{20pt}@{}p{226pt}@{}}
  \textbf{F1:} & {The most negative eigenvalue is obtained for ${M=d}$, and is
    equal to ${\frac{1}{d}-\half}$, the negative eigenvalue of
    $\widehat{\varrho}^{(d)}_{\textsc{ulin}}$ in Eq.~\eqref{eq:dit-1}.
    These values are marked by filled squares in Fig.~\ref{fig:1}.} \\[1ex]

\textbf{F2:} & {For ${2\leq M\leq d}$, there are always more $M$ values
    for which ``$>$'' applies in Eq.~\eqref{eq:neg-1} than those for
    which ``$=$'' is the case.  These values are marked by empty or
    filled circles, respectively.}\\[1ex]

\textbf{F3:} & { Some deviations from a monotonic decrease of
  $\lambda_{\text{min}}$ as $M$ increases from $1$ to $d$ are observed for
  some dimensions, notwithstanding the obvious trend.}
\end{tabular}

We leave it as a moot point whether or not these features are also
present in dimensions higher than ${d=13}$.  Should feature F1 be
generally true, then
\begin{equation}
  \label{eq:neg-5}
  \lambda_{\text{min}} \geq-\min\left\{\frac{M-1}{d},\half-\frac{1}{d}\right\}
\end{equation}
would sharpen the inequality in Eq.~\eqref{eq:neg-1}.  Currently, this
is just a conjecture suggested by the evidence presented in
Fig.~\ref{fig:1}.  Whether it holds for all prime-power dimensions
$d$, is perhaps of some interest for those who study the properties of
MUB.  It is, however, of no consequence for quantum state estimation,
where the fact that $\lambda_{\text{min}}$ is negative for all
intermediate $M$ values matters, while the precise value does not.

\section{Least-bias MUB inference}
\label{sec:least-bias}

\subsection{Physical unbiased estimators and von Neumann entropy}
\label{sec:vNent}

As we know, if the ULIN estimator
$\widehat{\varrho}^{(M)}_{\textsc{ulin}}$ is a \textsl{bona fide}
density operator, it maximizes the Shannon entropy of the unmeasured
probabilities in Eq.~\eqref{eq:bm-6}.  For ${d>2}$, only the
estimators for $M=1$ and $M=d+1$ will surely also yield the largest
von Neumann entropy of Eq.~\eqref{eq:bit-2}.  The permissible ULIN
estimators for ${1<M<d+1}$ do not generally maximize the von Neumann
entropy over the convex set of the physical
$\widehat{\varrho}^{(M)}_{\text{other}}$s of Eq.~\eqref{eq:bm-5a}:
they are not the estimators of Ref.~\cite{Teo:2011mn}. 

A simple counterexample for ${d=3}$ and ${M=2}$ is enough to support
this statement.  We return to Eq.~\eqref{eq:trit-5} and consider the
equal-weight case of ${\lambda_1=\lambda_2=\half}$, so that
${z_1=\half q^j}$, ${z_2=\half q^k}$, and ${z_3=z_4=0}$.  The rank-2
operators $\widehat{\varrho}=\widehat{\varrho}_{\textsc{ulin}}^{(2)}%
=\widehat{\varrho}_{\textsc{ulin}}^{(3)}=\widehat{\varrho}_{\textsc{ulin}}^{(4)}$
have eigenvalues $0,\frac{1}{6}(3\pm\sqrt{3})$ and von Neumann entropy
$S (\widehat{\varrho}^{(2)}_{\textsc{ulin}})=0.5157$.  The
largest value of the von Neumann entropy is $0.6370$, however; we
obtain it for the $\widehat{\varrho}^{(2)}_{\text{other}}$ with
$q^{j+k}\widehat{z}^{(2)}_3=q^{j-k+1}\widehat{z}^{(2)}_4=-0.09466$.

It is, of course, hardly surprising that the Shannon and the von
Neumann entropy are usually maximized by different
$\widehat{\varrho}^{(M)}_{\text{other}}$s.  These are two different
figures of merit, which serve different purposes: if we intend to
measure the remaining bases of the set of MUB, the Shannon entropy is
a useful quantity; by contrast, if we want to have a more universally
applicable estimator, then maximizing the von Neumann entropy is a
time-honored approach in Jaynes's
spirit~\cite{Jaynes:1957zl,Jaynes:1957rw,Buzek:2004fv}.  That the two
procedures coincide in the qubit case is nothing more than a result of
the geometrical simplicity of the problem.

\subsection{Physical estimators with least bias}
\label{sec:LBphilo}

Yet, in the context of incomplete MUB tomography, it is clearly more
natural to choose the ``unmeasured'' contribution in
Eq.~\eqref{eq:bm-5a} such that the Shannon entropy is maximized rather
than the von Neumann entropy.  For, the maximization of the Shannon
entropy ensures that we accept
$\widehat{\varrho}^{(M)}_{\textsc{ulin}}$ as the estimator whenever
that is permissible---that is, whenever
${\widehat{\varrho}^{(M)}_{\textsc{ulin}}\geq0}$---and when that is
not the case, we stay as close to naive unbiasedness as possible (more
about this closeness in Sec.~\ref{sec:LBcomp}).  In other words, it is
not necessary to completely discard the notion of unbiasedness; a
certain refinement is called for: we opt for the \emph{least-bias}
estimator $\widehat{\varrho}^{(M)}_{\textsc{lb}}$, which is the
physical $\widehat{\varrho}^{(M)}_{\text{other}}$ with the largest
Shannon entropy for the unmeasured probabilities,
\begin{equation}
\label{eq:lb-1}
  \max_{\widehat{\varrho}^{(M)}_{\text{other}}\geq0}
  \Eta^{(M)} ( \widehat{\varrho}^{(M)}_{\text{other}} )
  =\Eta^{(M)}  ( \widehat{\varrho}^{(M)}_{\textsc{lb}} ) \,.
\end{equation}
We repeat, perhaps unnecessarily, that the least-bias estimator is
different from the ULIN estimator only if the latter is unphysical.

\subsection{An efficient algorithm}
\label{sec:LBalgo}

The numerical search for $\widehat{\varrho}^{(M)}_{\textsc{lb}}$
requires the maximization of $\Eta^{(M)}(\varrho)$ over the convex set
of physical $\widehat{\varrho}^{(M)}_{\text{other}}$s; that is: over
the set of all density operators $\varrho$ with
${p_{\alpha k}=\tr(\varrho\Pi_{\alpha k})}$ for
${\alpha=1,2,\dots,M}$.  Other than that, $\varrho$ is only
constrained by the defining properties of a density operator:
${\varrho\geq0}$ and ${\tr\varrho=1}$.  This optimization problem is
very similar to the one solved in Ref.~\cite{Teo:2011mn}, where one
maximizes the von Neumann entropy over the convex set of
maximum-likelihood estimators.  Indeed, the algorithm of
Ref.~\cite{Teo:2011mn} can be modified so that it applies to the
current problem of finding $\widehat{\varrho}^{(M)}_{\textsc{lb}}$.

The figure of merit is the following function of $\varrho$:
\begin{eqnarray}
\label{eq:lb-2}
  D_{\mu}(\varrho) &=&
    \sum_{\alpha=1}^M\sum_{k=0}^{d-1}p_{\alpha k}
   \ln[ \tr(\varrho\Pi_{\alpha k})] 
\nonumber  \\ 
&  - & \mu\sum_{\beta=M+1}^{d+1}\sum_{l=0}^{d-1}
   \tr(\varrho\Pi_{\beta l})
 \ln[ \tr(\varrho\Pi_{\beta l} )]\,.\qquad
\end{eqnarray}
The first term is a mock log-likelihood for relative frequencies equal
to the known probabilities $p_{\alpha k}$ for ${\alpha\leq M}$.  The
second term is the Shannon entropy $\Eta^{(M)}(\varrho)$ of multiplied
by the nonnegative parameter $\mu$.

Now, all physical $\widehat{\varrho}^{(M)}_{\text{other}}$s 
maximize  the ${\mu=0}$ function $D_0(\varrho)$, with
\begin{equation}
\label{eq:lb-2'}
\max_{\varrho}D_0(\varrho) = 
D_0 ( \widehat{\varrho}^{(M)}_{\text{other}} ) 
=\sum_{\alpha=1}^M\sum_{k=0}^{d-1} p_{\alpha k}\ln p_{\alpha k}\,.
\end{equation}
Because of this extremal property of
$D_0 ( \widehat{\varrho}^{(M)}_{\text{other}} )$, the first-order
contribution to $D_{\mu}(\varrho)$ originates solely in the
Shannon-entropy term,
\begin{equation}\label{eq:lb-3}
  \max_{\varrho}D_{\mu}(\varrho)
  =D_0 (\widehat{\varrho}^{(M)}_{\text{other}} )
  +\mu \Eta^{(M)} ( \widehat{\varrho}^{(M)}_{\textsc{lb}}  ) + \cdots\,,
\end{equation}
where the ellipsis stands for contributions of order $\mu^2$ and
higher.  Therefore, the numerical maximization of $D_{\mu}(\varrho)$
for sufficiently small $\mu$ yields the least-bias estimator
$\widehat{\varrho}^{(M)}_{\textsc{lb}}$.  Practical experience shows
that ${\mu\approx10^{-4}}$ is large enough for a noticeable difference
between $D_{\mu}(\varrho)$ and $D_0(\varrho)$ and small enough to
ensure that the density operator that maximizes $D_{\mu}(\varrho)$
approximates $\widehat{\varrho}^{(M)}_{\textsc{lb}}$ very well.
Usually, it is a good idea to numerically select a $\mu$ that
minimizes the fluctuations in the extremal $\varrho$---from the
iteration described below---to some pre-chosen precision.

The first-order response of $D_{\mu}(\varrho)$ to a variation
$\varrho \mapsto \varrho+\updelta\varrho$ is
\begin{equation}
\label{eq:lb-4}
  \updelta D_{\mu}(\varrho)= \tr[\updelta\varrho\,W(\varrho)] \, ,
\end{equation}
with
\begin{eqnarray}
\label{eq:lb-5}
 W(\varrho) & = & \sum^M_{\alpha=1}\sum^{d-1}_{k=0}
\dfrac{p_{\alpha k}}{\tr(\varrho\Pi_{\alpha k})}\Pi_{\alpha k} \nonumber\\
& - & {\mu\sum^{d+1}_{\beta=M+1}\sum^{d-1}_{l=0} 
\Pi_{\beta l} \ln[ \tr(\varrho\Pi_{\beta l})]} \,.
\end{eqnarray}
Hence, the positivity constraint for $\varrho$ implies the 
following extremal equation for $\widehat{\varrho}^{(M)}_\textsc{lb}$:
\begin{equation}
\label{eq:lb-6}
W (\widehat{\varrho}^{(M)}_\textsc{lb} ) 
 \widehat{\varrho}^{(M)}_\textsc{lb}
= \tr [ W (\widehat{\varrho}^{(M)}_\textsc{lb} )
\widehat{\varrho}^{(M)}_\textsc{lb} ]
\widehat{\varrho}^{(M)}_\textsc{lb} \, .
\end{equation}

Since $D_{\mu}(\varrho)$ has no local maxima, we can find
$\widehat{\varrho}^{(M)}_\textsc{lb}$ by steepest ascent (``follow the
gradient uphill'').  The $W(\varrho)$ identifies the gradient in the
sense that
\begin{equation}
\label{eq:lb-7}
  \updelta \varrho \propto  W(\varrho) \varrho +
\varrho W(\varrho) -2\varrho\tr[\varrho W(\varrho)]
\end{equation}
gives the largest first-order change in $D_{\mu}(\varrho)$.
We ensure a positive increment of $D_{\mu}(\varrho)$ by the iteration in
accordance with
\begin{widetext}
\begin{equation}
\label{eq:lb-8}
\varrho_n \mapsto\varrho_{n+1}
= \dfrac{\bigl(\openone + \epsilon \bigl\{ W(\varrho_n) -
  \tr [ W(\varrho_n)\varrho_n]\openone\bigr\}\bigr)  \,
  \varrho_n
 \bigl(\openone+\epsilon \bigl\{ W(\varrho_n)
 - \tr [ W(\varrho_n)\varrho_n]\openone \bigr\}  \bigr)}
   {1+\epsilon^2\tr\Bigl(\bigl\{ W(\varrho_n) -
  \tr [ W(\varrho_n)\varrho_n]\openone \bigr\}^2\varrho_n\Bigr)}\,,
\end{equation}
\end{widetext}
where $\epsilon$ is a small positive step size.  To obtain the
extremal solution $\widehat{\varrho}^{(M)}_\textsc{lb}$, one can start
with the maximally-mixed state ${\varrho_{n=1}=\frac{1}{d}\openone}$
and iterate Eq.~\eqref{eq:lb-8} for a small $\mu$ until the extremal
equations (\ref{eq:lb-6}) are obeyed to a satisfactory accuracy.

We note, but do not elaborate this point, that one can speed up the
convergence substantially by employing conjugate
gradients~\cite{Shewchuk:1994oz}; each iteration step is then more
costly (in CPU time) but that is more than compensated for by the much
smaller number of steps.  Further we note that, in cases where the
iteration algorithm proceeds too slowly due to the complexity of the
optimization problem for large Hilbert-space dimensions, the sum of
the two entropic functions in Eq.~\eqref{eq:lb-2} can alternatively be
optimized with gradient-free methods, such as the Nelder--Mead
``amoeba'' method~\cite{Nelder:1965rz} or simulated
annealing~\cite{vanLaarhoven:1985ai}.

\subsection{Examples}
\label{sec:LBexpl}

\begin{table}[b]
\caption{\label{tbl:ULIN-LB}%
Comparison between the ULIN and least-bias estimators that maximize
the Shannon entropy for $\varrho_w$.  When $\varrho_w$ is the pure
state  defined by Eq.~\eqref{eq:trit-6} ($w=0$), the least-bias
estimator is the true state $\varrho$. Also, note that for
$\varrho_w$, $z_{1} = z_{2} = z_{3} = -(1-w)/2$, $z_{4} = -(1-w) q^{2}/2$.}
\begin{ruledtabular}
\begin{tabular}{lcccc}
& \multicolumn{2}{c}{$w=0.1$} \hspace{0.5cm} & \multicolumn{2}{c}{$w=0.2$}  \\
  \cline{2-3}   \cline{4-5}  
&   $z_3$ & $z_4$ &  $z_3$ & $z_4$ \\ \hline
 $\varrho_w$ & $-0.450$ &  $0.225 + \I\,0.389$ &
 $-0.400$ &  $0.200 + \I\,0.346$ \\[0.5ex]
  $\widehat{\varrho}^{(2)}_\textsc{ulin}$  & 0 & 0 & $0$ & $0$\\
    $\widehat{\varrho}^{(2)}_\textsc{lb}$ & $-0.313$ 
 & $0.156+\I\,0.271$ & $-0.126$ & $0.063+\I\,0.109$\\[0.5ex]
    $\widehat{\varrho}^{(3)}_\textsc{ulin}$ & $-0.450$ & 0
   & $-0.400$ & $0$\\
    $\widehat{\varrho}^{(3)}_\textsc{lb}$  & $-0.450$ &
    $0.174+\I\,0.303$ & $-0.400$ & $0.100+\I\,0.173$ 
\end{tabular}
\end{ruledtabular}
\end{table}

For explicit examples, we look at a class of qutrit states comprising
statistical mixtures of $\varrho$ as represented in \eqref{eq:trit-6}
with the maximally-mixed state for various admixtures $0\leq w\leq1$:
\begin{equation}
  \varrho \mapsto \varrho_w=(1-w)\varrho+w\dfrac{\openone}{3}\,.
\end{equation}
As always, for $M=1$ and $M=4$, the ULIN estimator is always positive
semidefinite. Beyond ${w=0.2679}$ and ${w=\third}$ respectively for
$M=2$ and $M=3$, the ULIN estimator is precisely the physical
least-bias estimator that maximizes the Shannon entropy since the true
state would be highly mixed. For other $w$ values, the ULIN estimator
possesses at least one negative eigenvalue. The relevant parameters
for the unphysical ULIN and physical least-bias estimators for two
exemplifying values of $w$ are listed in Table~\ref{tbl:ULIN-LB}.

\section{Comparison with other estimators}
\label{sec:LBcomp}

\subsection{Estimators of the least-bias kind}
\label{sec:LBcomp-1}

The Shannon entropy quantifies our ignorance about the outcomes of
future projective measurements in the yet-unmeasured bases with
${\beta=M+1,M+2,\dots,d+1}$ from the set of MUB.  Rather than
ignorance we can equivalently quantify knowledge by the
\emph{predictability} of the future measurements.  For an experiment
with probabilities $p_0,p_1,\dots,p_{d-1}$, the entropic measure of
predictability is
\begin{equation}
\label{eq:oth-1}
  P_{\text{ent}}(p_{\cdot})=\sum_{l=0}^{d-1}p_l\log_d(p_l d)\, ,
\end{equation}
where the symbol $p_{\cdot}$ stands for all the $d$ probabilities. 
We have the extreme values
\begin{equation}
\label{eq:oth-2}
 P_{\text{ent}}(p_{\cdot})=\left\{
   \begin{array}{c@{\qquad \text{if}\ }l}
     0 & \ds p_0=p_1=\cdots=p_{d-1}=\frac{1}{d}\,,\\[1.5ex]
     1 & \ds p_l=\delta_{l\bar{l}}\quad 
   \mbox{for a certain $\bar{l}$}\,,
   \end{array}\right.
\end{equation}
and ${0<P_{\text{ent}}(p_{\cdot})<1}$ for all other sets of probabilities,
which are among the defining properties of all permissible measures of
predictability~\cite{Durr:2001bv,Englert:2008oe}. Convexity
\begin{equation}
 \label{eq:oth-2'}
  P(\lambda p_{\cdot}+\lambda^{\prime} p^{\prime}_{\cdot})\leq
 \lambda P(p_{\cdot})+ \lambda^{\prime} P(p^{\prime}_{\cdot} )
\end{equation}
for ${0\leq\lambda=1-\lambda'\leq1}$ is another important property of all
predictability measures.

\begin{figure}[t]
  \centering
  \includegraphics[width=0.95 \columnwidth]{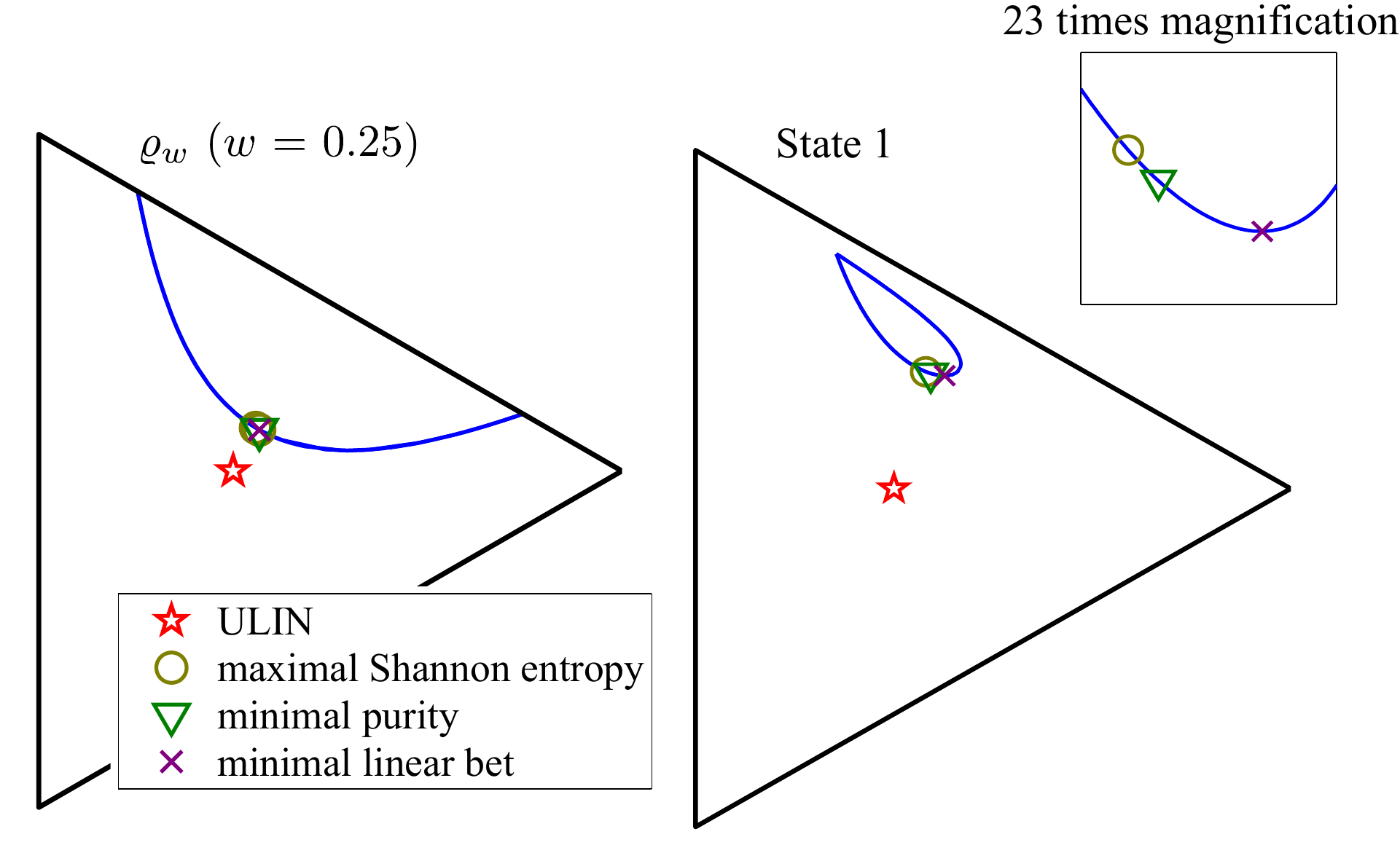}
  \caption{\label{fig:2} A prototypical showcase of three types of
    physical least-bias estimators relative to the ULIN estimator for
    different qutrit true states (${M=3}$).  The black triangle is the
    regular simplex for $z_4$ in the complex plane, with its corners
    at ${z_4=1}$, $q$, and $q^2$.  The blue curve borders the regions
    of permissible values for $z_4$.  The examples featured here are
    $\varrho_w$ for ${w=\frac{1}{4}}$ and state~1 of
    Table~\ref{tbl:MSE-LVNE-BME-LME}.  The magnified view is for
    state~1, where the three least-bias estimators are different; by
    contrast, all three are the same for $\varrho_w$.}
\end{figure}

In terms of $P_{\text{ent}}(p_{\cdot})$, the Shannon entropy is
\begin{equation}
\label{eq:oth-3}
  \Eta^{(M)}(\varrho)=\left [ d+1-M-
  \sum_{\beta=M+1}^{d+1} \!P_{\text{ent}} (
  \tr(\varrho\Pi_{\beta\cdot}) ) \right ] \ln d\,,
\end{equation}
so that we could replace the figure of merit in Eq.~\eqref{eq:lb-2} by
\begin{eqnarray}
\label{eq:oth-4}
  D_{\mu}(\varrho) &=&
    \sum_{\alpha=1}^M\sum_{k=0}^{d-1} p_{\alpha k}
 \ln [ \tr(\varrho\Pi_{\alpha k})]
\nonumber\\  
 & - & \mu\sum_{\beta=M+1}^{d+1}\! 
 P_{\text{ent}} ( \tr(\varrho\Pi_{\beta\cdot}) )\,,
\end{eqnarray}
with a corresponding minor change in Eq.~\eqref{eq:lb-5}, 
namely ${\mu \mapsto\mu/\ln d}$.

Other measures of predictability could be employed as well. For
example, the purity-based predictability
\begin{equation}
\label{eq:oth-5}
  P_{\text{pur}}(p_{\cdot})
   =\frac{d}{d-1}\sum_{l=0}^{d-1} \left ( 
 p_l-\frac{1}{d} \right )^2 \,.
\end{equation}
Upon replacing $P_{\text{ent}}(\ )$ in \eqref{eq:oth-4} by
$P_{\text{pur}}(\ )$, the $\widehat{\varrho}^{(M)}_{\text{other}}$
that maximizes $D_\mu(\varrho)$ is simply another least-bias estimator
$\widehat{\varrho}^{(M)}_{\textsc{lb}}$ than the one for
$P_{\text{ent}}(\ )$.  The two least-bias estimators are different
since they refer to different quantifications of the bias regarding
the probabilities for the not-yet-measured bases. But neither
$\widehat{\varrho}^{(M)}_{\textsc{lb}}$ is better than the other; they
serve different purposes.  Any other predictability $P(p_{\cdot})$
yields a corresponding least-bias estimator.  Our preference, in
Sec.~\ref{sec:least-bias}, for the
$\widehat{\varrho}^{(M)}_{\textsc{lb}}$ associated with
$P_{\text{ent}}(p_{\cdot})$ is mostly for historical reasons---history
of the subject, that is. 

Yet another example is provided by the predictability functions that
equal the expected gain for various strategies of betting on the
future outcome.  There is, in particular, the ``linear bet'' of
Sec.~2.3.1 in Ref.~\cite{Englert:2008oe}, which amounts to
\begin{equation}
\label{eq:linbet}
  P_{\text{bet}} (p_{\cdot} ) = 
  \max \{p_0,p_1,p_2\} - \min \{p_0,p_1,p_2\}
\end{equation}
in the qutrit case.

We can regard $P_{\text{pur}}(p_{\cdot})$ in \eqref{eq:oth-5} as the
normalized squared Hilbert-Schmidt distance of $p_0,p_1,\dots,p_{d-1}$
from the uniform distribution ${p_l=\frac{1}{d}}$.  Likewise,
$P_{\text{ent}}(p_{\cdot})$ is the normalized Kullback-Leibler
divergence of the actual probabilities from the uniform distribution.
Analogous remarks apply to any other permissible predictability
$P(p_{\cdot})$.  It follows that, irrespective of which $P(p_{\cdot})$
we choose, the resulting least-bias estimator equals the ULIN
estimator whenever ${\widehat{\varrho}^{(M)}_{\textsc{ulin}}\geq0}$.
And when ${\widehat{\varrho}^{(M)}_{\textsc{ulin}}\not\geq0}$,
then---as a consequence of \eqref{eq:oth-2'}---each
$\widehat{\varrho}^{(M)}_{\textsc{lb}}$ sits on the border of the
convex set of physical $\widehat{\varrho}^{(M)}_{\text{other}}$s,
whatever border point is closest to
$\widehat{\varrho}^{(M)}_{\textsc{ulin}}$ in the sense specified by
the choice of $P(p_{\cdot})$.

\begin{figure}[t]
  \centering
  \includegraphics[width=.65 \columnwidth]{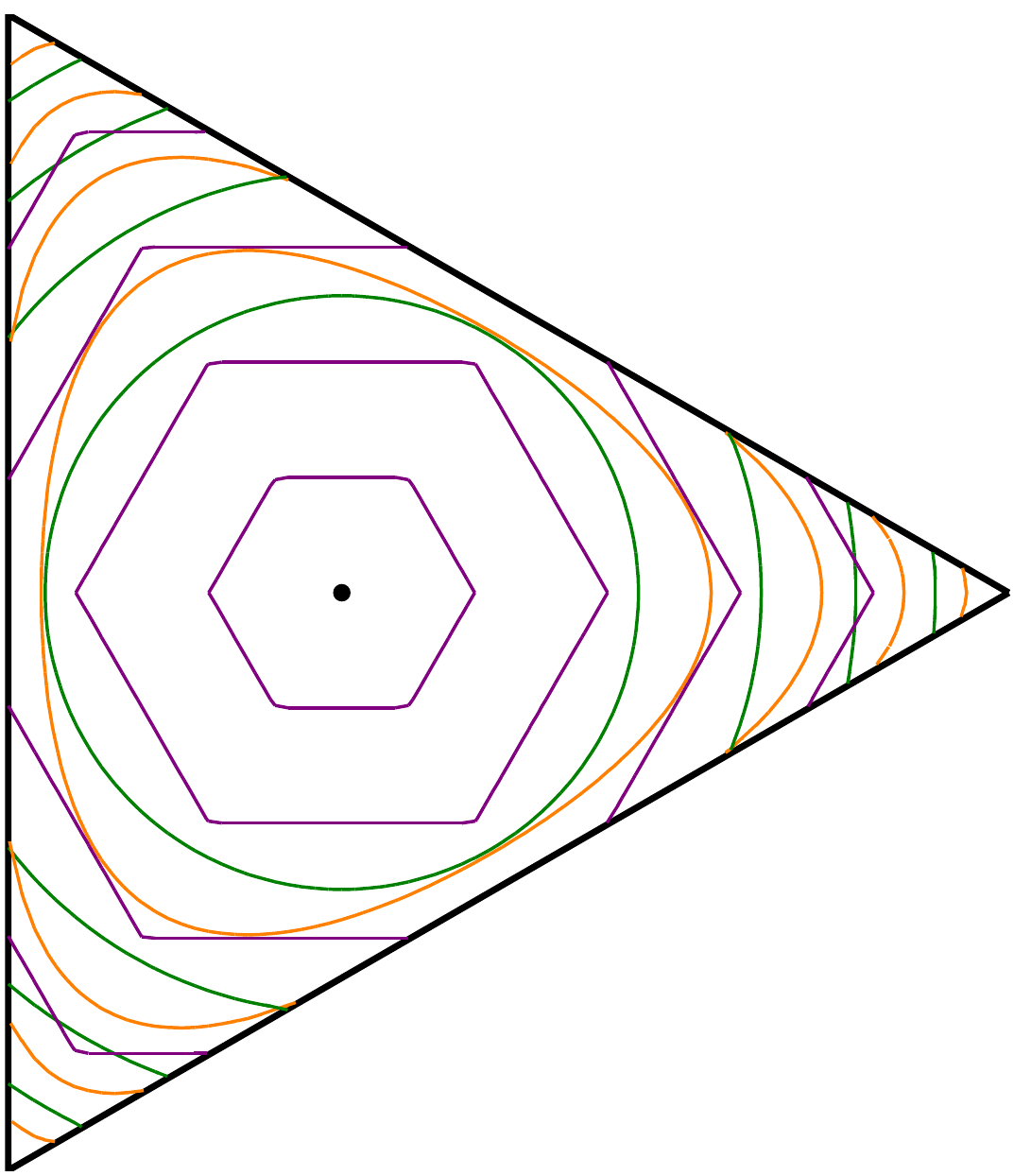}
  \caption{\label{fig:3} A comparison of the lines of constant
    ``distance'' from the ULIN estimator for the three different
    predictability functions that are used as figure of merit for the
    least-bias estimators in Figs.~2 and 4.  We have ${P=0}$ at the
    center of the triangle and ${P=1}$ in each corner, and the contour
    lines indicate predictability values ${P=0.2}$, $0.4$, $0.6$, and
    $0.8$, respectively.  The green circles are for the purity measure
    of Eq.~\eqref{eq:oth-5}; the orange distorted circles are for the
    Shannon-entropy measure of Eq.~\eqref{eq:oth-1} and the purple
    hexagons are for the betting measure of Eq.~\eqref{eq:linbet}.}
\end{figure}

These matters are illustrated in Fig.~\ref{fig:2}, which graphically
shows the situation for qutrit MUB where $M=3$ bases are measured. As
it turns out, numerical experience shows that different types of
least-bias estimators give estimators that are quite close to each
other, hinting that the choice of measure for quantifying bias
typically influences the resulting estimator only mildly as long as
these nonpathological choices lead to the same extremal solution in
the absence of the quantum positivity constraint. Such differences
would not matter much in the presence of statistical fluctuations and
other experimental error sources. For various qutrit states, the two
kinds of least-bias estimators introduced in this section are
essentially identical. For completeness, in Fig.~\ref{fig:3} we
present a comparison of the different figures of merit for the
least-bias kind.

\subsection{Other estimators}
\label{sec:LBcomp-2}

The least-bias estimators thus constructed are rank-deficient whenever
${\widehat{\varrho}^{(M)}_{\textsc{ulin}}\not\geq0}$.  It may,
however, be desirable to use estimators that usually have full
rank---they are more robust in the sense that a slight perturbation, or
inaccuracy in determining them, does not render them unphysical. 
Other reasons for avoiding rank-deficient estimators have been put
forward as well~\cite{Blume-Kohout:2010wc}.

From the plethora of physical
$\widehat{\varrho}^{(M)}_{\text{other}}$s, we consider three choices:
the estimator with the largest von Neumann entropy~\cite{Teo:2011mn},
the Bayesian mean estimator for a uniform prior on the unmeasured
probabilities~\cite{Blume-Kohout:2010ya}, and the estimator having the
largest minimal eigenvalue~\cite{Wilson:2014jb,Hedayat:1981sh}.

\begin{figure}[h]
  \centering
  \includegraphics[width=0.95 \columnwidth]{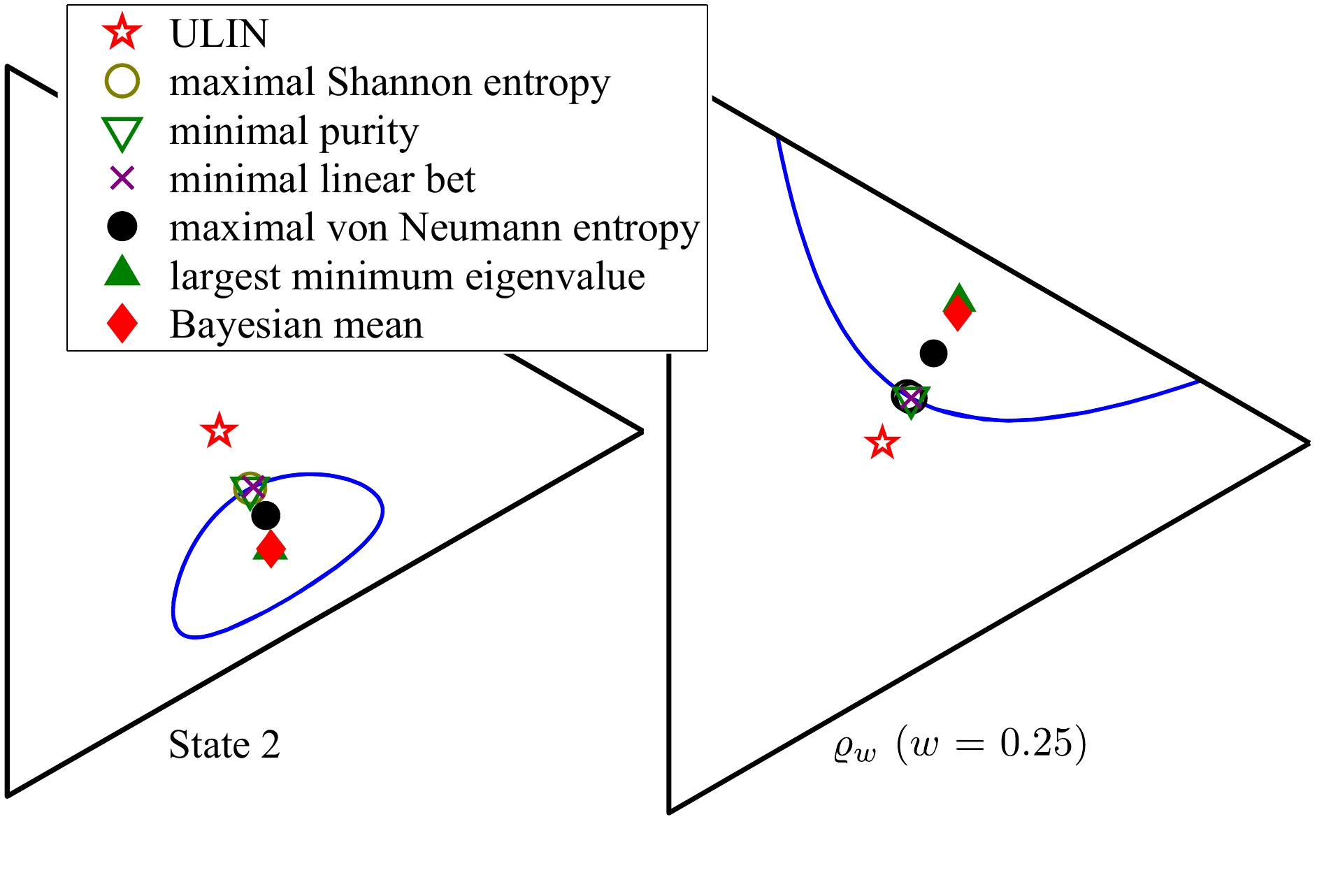}
  \caption{\label{fig:4} Pictorial comparison of the ULIN estimator
    with three physical full-rank estimators: the estimators with
    largest von Neumann entropy and largest minimal eigenvalue, and
    the Bayesian mean estimator.  The rank-deficient three least-bias
    estimators of Fig.~2 are also indicated; they sit on the blue
    border of the region of permissible states and are too close to
    each other for telling them apart.  As in Fig.~\ref{fig:2}, the examples are
    for qutrit states with ${M=3}$, and the triangle is the simplex
    for $z_4$.  The examples featured here are state~2 and $\varrho_w$
    for ${w=\frac{1}{4}}$ of Table~\ref{tbl:MSE-LVNE-BME-LME}.}
\end{figure}

These estimators simply differ in the unmeasured probabilities, as is
the case for any estimator consistent with the probabilities of the
already-measured bases. To streamline presentations, Fig.~\ref{fig:4}
and Table~\ref{tbl:MSE-LVNE-BME-LME} illustrate the estimators for the
$M=3$ case of a qutrit.

\section{Conclusions}
\label{sec:con}

In summary, we have examined whether the ideal Laplace-type linear
estimator that assigns equal and unbiased probabilities to all the
outcomes of unmeasured bases from the set of MUB is a physical thing
to do in incomplete MUB tomography.  The answer is definite: such
unbiased estimation does not work as a general recipe.

\begin{table}[t]
  \caption{\label{tbl:MSE-LVNE-BME-LME}
    Values of $z_4$ for physical estimators to three unphysical ULIN estimators
    for qutrit states with three measured bases (${d=M=3}$).
    These data are used in Figs.~\ref{fig:2} and \ref{fig:4}.
    The values of $z_1$, $z_2$, and $z_3$ that specify the respective 
    $\widehat{\varrho}_{\textsc{ulin}}^{(3)}$s are 
    (i) ${z_1=z_2=z_3=-3/8}$ for $\varrho_{w=1/4}$; 
    (ii) ${z_1=0.160-\I\,0.321}$, ${z_2=0.571-\I\,0.192}$, and
    ${z_3=0.314+\I\,0.165}$ for state~1; 
    and (iii) ${z_1=-0.345+\I\,0.0574}$, ${z_2=0.303+\I\,0.328}$, and
    ${z_3=0.00057-\I\,0.294}$ for state~2.
    The rows report the corresponding three $z_4$ values for the least-bias
    estimator $\widehat{\varrho}^{(3)}_\textsc{lb}$ of
    Secs.~\ref{sec:LBphilo} and \ref{sec:LBalgo} for the alternative
    least-bias  estimators of Sec.~\ref{sec:LBcomp-1} 
    $\widehat{\varrho}^{(3)}_\textsc{pur}$ and $\widehat{\varrho}^{(3)}_\textsc{bet}$;
    for the full-rank estimators with
    largest von Neumann entropy and largest minimal eigenvalue,
    $\widehat{\varrho}^{(3)}_\textsc{vN}$ and $\widehat{\varrho}^{(3)}_\textsc{mineig}$;
    and for the Bayesian mean estimator $\widehat{\varrho}^{(3)}_\textsc{bm}$.
    For information, the last row contains the $z_4$ values of the actual states 
    that were used to obtain the values of $z_1$, $z_2$, $z_3$ for the
    $\widehat{\varrho}_{\textsc{ulin}}^{(3)}$s.}
  \centering
  \begin{ruledtabular}
    \begin{tabular}{cccc}
      & $ \varrho_{w=\frac{1}{4}}$ & state~1 &  state~2  \\
      & $z_{4}$ &  $z_{4}$ & $z_{4}$ \\  \hline
      $\widehat{\varrho}^{(3)}_\textsc{lb}$ & $0.067+\I\,0.106$ 
      & $0.080+\I\,0.299$  & $0.073-\I\,0.136$ \\
      $\widehat{\varrho}^{(3)}_\textsc{pur}$ & $0.067+\I\,0.106$ 
      & $0.093+\I\,0.295$ & $0.073-\I\,0.136$ \\
      $\widehat{\varrho}^{(3)}_\textsc{bet}$ & $0.067+\I\,0.106$ 
      & $0.128+\I\,0.289$ & $0.080-\I\,0.132$ \\[0.5ex]
      $\widehat{\varrho}^{(3)}_\textsc{vN}$ & $0.120+\I\,0.208$ 
      & $0.090+\I\,0.309$  & $0.104-\I\,0.204$ \\
      $\widehat{\varrho}^{(3)}_\textsc{mineig}$ & $0.187+\I\,0.325$ 
       & $0.003+\I\,0.438$ & $0.122-\I\,0.283$ \\
      $\widehat{\varrho}^{(3)}_\textsc{bm}$ & $0.176+\I\,0.305$ 
       & $0.021+\I\,0.418$ & $0.122-\I\,0.279$ \\[0.5ex]
      actual state & $\frac{º3}{16}+\I\,0.324$ 
       & $-0.146+\I\,0.601$  & $0.256-\I\,0.337$  \\[0.3ex]
    \end{tabular}
  \end{ruledtabular}
\end{table}

As a natural adjustment to the original inference method that follows
a blind application of Laplace's notion of indifference, we recommend
the use of the least-bias estimator.  It gives the closest-to-uniform
unmeasured probability distributions by maximizing the Shannon entropy
with due attention to the constraints imposed by the measured
probabilities.  We also supply a simple iterative algorithm for
computing the least-bias estimator.

We compared this least-bias estimator with alternative estimators that
quantify the bias differently, and concluded that these alternatives
are equally useful for practical purposes.  Since all these least-bias
estimators are rank-deficient whenever they are different from the
linear estimator of ideal Laplace-type, we also took a look at three
different estimators of full rank.  As illustrated by examples, the
full-rank estimators are indeed different from the least-bias
estimators, but are certainly acceptable as consistent \textsl{bona
  fide} estimators.

Armed with these insights, one could now study questions such as
\textit{after measuring in $M$ bases, which basis is optimal for the
  next von Neumann test?}  This matter and others are, however, beyond
the scope of the present article.  

\begin{acknowledgments}
  Many of the ideas in this paper originated at the Workshop on
  Mathematical Methods of Quantum Tomography at Fields Institute
  (Toronto) in February 2013.  Z.~H., J.~R., and Y.~S.~T.  are
  grateful for the support of the European Social Fund and the state
  budget of the Czech Republic [project No. CZ.1.07/2.3.00/30.0004
  (POST-UP)], the Grant Agency of the Czech Republic (Grant
  No. 15-031945), the IGA Project of the Palack{\'y} University (Grant
  No. IGA PrF 2015-002), and the sustainability of Post-Doc positions
  at Palack{\'y} University.  L.~L.~S.~S. acknowledges the support
  from UCM-Banco Santander Program (Grant GR3/14) and helpful
  discussions with Markus Grassl. H.~K.~N.'s, J.~H.~C.'s, and
  B.-G.~E.'s work is funded by the Singapore Ministry of Education
  (partly through the Academic Research Fund Tier 3 MOE2012-T3-1-009)
  and the National Research Foundation of Singapore.  H.~K.~N is also
  funded by a Yale-NUS College start-up grant.
\end{acknowledgments}


%

\end{document}